\documentclass[twocolumn,showpacs,preprintnumbers,amsmath,amssymb, prb]{revtex4-1}

\usepackage{stmaryrd}
\usepackage{txfonts}
\usepackage{amssymb}
\usepackage{mathrsfs}
\usepackage{graphicx}
\usepackage{dcolumn}
\usepackage{bm}
\usepackage{epsfig}

\begin{document}
\preprint{S.-Z. Lin and L. N. Bulaevskii, Phys. Rev. B {\bf{86}}, 014518 (2012)}

\title{Theory for measurements of penetration depth in magnetic superconductors by magnetic force microscopy and scanning SQUID microscopy}

\author{Shi-Zeng Lin and Lev N. Bulaevskii}
\affiliation{Theoretical Division, Los Alamos National Laboratory, Los Alamos, New Mexico 87545}

\begin{abstract}
The working principle of magnetic force microscopy and scanning SQUID microscopy is introducing a magnetic source near a superconductor and measuring the magnetic field distribution near the superconductor, from which one can obtain the penetration depth. We investigate the magnetic field distribution near the surface of a magnetic superconductor when a magnetic source is placed close to the superconductor, which can be used to extract both the penetration depth $\lambda_L$ and magnetic susceptibility $\chi$ by magnetic force microscopy or scanning SQUID microscopy. When the magnetic moments are parallel to the surface, one extracts $\lambda_L/\sqrt{1-4\pi \chi}$. When the moments are perpendicular to the surface, one obtains $\lambda_L$. By changing the orientation of the crystal, one thus is able to extract both $\chi$ and $\lambda_L$.
 \end{abstract}
\date{\today}
\maketitle

\section{Introduction}
Superconductivity is well characterized by two length scales. The coherence length $\xi$ describes the rigidity of the phase coherence and the penetration depth $\lambda_L$ characterizes the response to electromagnetic fields. The penetration depth is directly connected to the superfluid density and the pairing symmetry, thus its measurement is crucial for the understanding of new discovered superconductors. The coherence length can be measured from the upper critical field $H_{c2}$ for type II superconductors. There are many well-developed method to measure the penetration depth\cite{Prozorov2006}, such as the magnetic force microscopy (MFM) and scanning SQUID microscopy (SSM), which are the main focus of the present study.   

In MFM and SSM, a magnetic source is placed near a superconductor\cite{Kirtley1996,Kirtley1999,Kirtley2010}. In MFM, the source magnetic field is generated by a small magnetic tip, which can be modeled as a point dipole. In SSM, the source field is generated by a current loop. Due to the exclusion of magnetic field by the superconductor, the magnetic field outside the superconductor is modified compared to that without the superconductor. The exclusion thus causes repulsion between the magnetic source and superconductor. In MFM, the resulting magnetic field distribution is measured by the force between the MFM tip and superconductor. In SSM, the magnetic field is measured by a SQUID. From the measured magnetic field, one can extract the penetration depth by fitting to theoretical expressions. For non-magnetic superconductors, the magnetic field distribution was calculated in Refs. \onlinecite{Xu1995,Coffey1995,Coffey1998,Coffey1999,Badia2001}  for isotropic superconductors and in Ref. \onlinecite{Kogan2003} for anistropic superconductors.

Recently there is growing interest to apply both MFM and SSM to magnetic superconductors, where magnetic ordering coexists with superconductivity\cite{Bulaevskii85,Buzdin05, Gupta2006}. When a magnetic field induced by a source is applied to the magnetic superconductors, it polarizes the magnetic moments near the surface of the superconductors, which gives additional contribution to the magnetic field outside the superconductors. The polarization is characterized by the magnetic susceptibility $\chi$ in the linear response approximation. For instance, in MFM, the polarization lowers the energy of the whole system, thus gives attraction contribution between the MFM tip and superconductor in additional to repulsion due to the screening of magnetic field by superconductors. The magnetic field distribution outside the magnetic superconductor thus depends on $\chi$ and $\lambda_L$. It is still an open question what information can be extracted by MFM and SSM in the case of magnetic superconductors. Recently Kirtley \emph{et. al.} studied the SSM response in isotropic paramagnetic superconductors. \cite{Kirtley2012} The effects of the isotropic paramagnet in this case are two folds. First it reduces the penetration depth according to $\lambda_L\sqrt{1-4\pi\chi}$. Second, it changes the boundary condition. In isotropic paramagnetic superconductors, the magnetic field outside depends on $\lambda_L/\sqrt{1-4\pi \chi}$, and one cannot extract both $\lambda_L$ and $\chi$ from SSM measurements.

The magnetic superconductors usually have anisotropy in magnetic structure. The polarization depends on the orientation of the magnetic source with respect to the anisotropy of the magnetic structure. By changing the orientation of the crystal, it is possible to obtain both $\lambda_L$ and $\chi$.  

In this work, we investigate the magnetic response in magnetic superconductors based on the London approach both in the Meissner state and mixed state.  For the magnetic moments parallel to the surface of superconductor, the magnetic field outside the superconductor depends on $\lambda_L/\sqrt{1-4\pi \chi}$ when the separation between the magnetic source and superconductor is much larger than $\lambda_L$. For the moments perpendicular to the surface, it depends only on $\lambda_L$. By changing the orientation of the crystal, one thus can obtain both the bare penetration depth and the magnetic susceptibility.

\section{Model}
\begin{figure}[t]
\psfig{figure=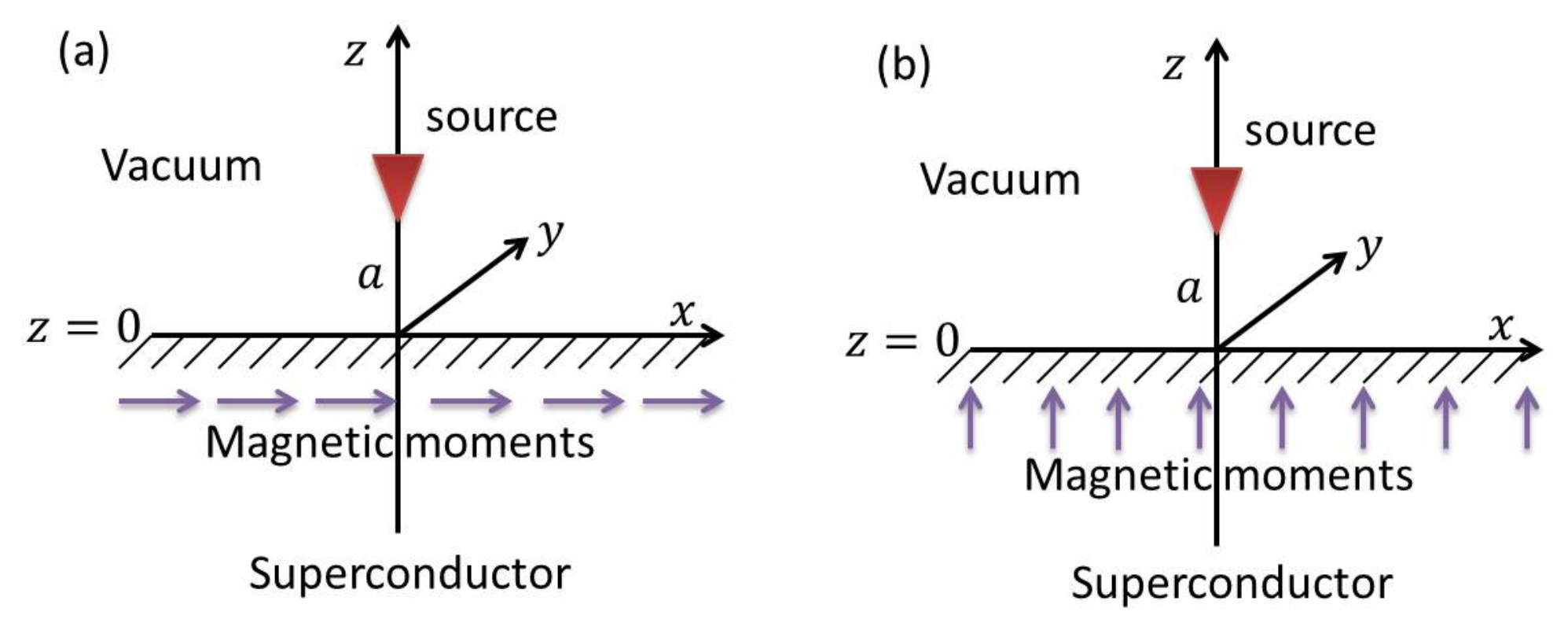,width=\columnwidth}
\caption{\label{f1}(color online) Schematic view of a magnetic source placed on top of a magnetic superconductor. The magnetic subsystem has easy-axis anisotropy. The easy axis is parallel to the surface in (a) or perpendicular to the surface in (b). } 
\end{figure}

In this section, we derive the magnetic field distribution inside and outside the magnetic superconductor. A schematic view of the setup is shown in Fig. \ref{f1}. To be specific, we consider the case with an easy-axis anisotropy in magnetic structure, which is most commonly encountered in magnetic superconductors\cite{Bulaevskii85,Gupta2006}. The penetration depth is assumed to be isotropic. We consider two cases with magnetic moment parallel to the surface Fig. \ref{f1}(a) and perpendicular to the surface Fig. \ref{f1}(b).

The magnetic field outside the superconductor is given by
\begin{equation}\label{eq1}
\nabla\times(\mathbf{B}-4\pi \mathbf{M_s})=0,
\end{equation}
where $\mathbf{M}_s$ is the magnetization in the source. Using $\nabla\cdot \mathbf{B}=0$ we can rewrite Eq. (\ref{eq1}) as 
\begin{equation}\label{eq2}
\nabla^2 \mathbf{B}=-4\pi\nabla\times\nabla\times \mathbf{M}_s.
\end{equation}
Inside the superconductor, we use the London approximation which is valid when the penetration length is much larger than the coherence length as realized in most magnetic superconductors\cite{Tachiki79,Gray83,Buzdin84,Bulaevskii85}
\begin{equation}\label{eq3}
\lambda_L^2\nabla\times\nabla\times(\mathbf{B}-4\pi \mathbf{M})+\mathbf{B}=\Phi_0\delta(x)\delta(y)\hat{\mathbf{z}},
\end{equation}
where $\hat{\mathbf{z}}$ is the unit vector along the $z$ axis and $\Phi_0=hc/(2e)$ is the flux quantum. We assume the vortex density is small when the applied magnetic field is much smaller than $H_{c2}$, and we only consider a single vortex at $(x, y)=(0, 0)$.

The magnetic field outside and inside the superconductor is connected through the boundary conditions at the interface $z=0$. The normal component of $\mathbf{B}$ is continuous at the interface
\begin{equation}\label{eq4}
B_z(z=0^-)=B_z(z=0^+).
\end{equation}
As the induced surface supercurrent is finite at the interface, from the Maxwell equations we have the boundary condition for the tangential component $B_{x,y}$
\begin{equation}\label{eq5}
\left( {\begin{array}{cc}
 B_{x}(z=0^-)  \\
 B_{y}(z=0^-)
 \end{array} } \right)-4\pi \left( {\begin{array}{cc}
 M_{x}(z=0^-)  \\
 M_{y}(z=0^-)
 \end{array} } \right)=\left( {\begin{array}{cc}
 B_{x}(z=0^+)  \\
 B_{y}(z=0^+)
 \end{array} } \right).
\end{equation}

The general solution to Eq. (\ref{eq2}) can be written as $\mathbf{B}=\mathbf{B}_1+ \mathbf{B}_2$, where $\mathbf{B}_1$ and $\mathbf{B}_2$ are solutions of Eq. (\ref{eq2}) with $\mathbf{M}_s$ and without $\mathbf{M}_s$ respectively. $\mathbf{B}_2(\mathbf{r})$ can be written as
\begin{equation}\label{eq6}
\mathbf{B}_2(\mathbf{r})=\frac{1}{(2\pi )^{3/2}}\int dk_xdk_y \mathbf{B}_2(k) \exp \left[i \mathbf{k}_{2d}\cdot\mathbf{r}_{2d}-k_{2d}z \right],
\end{equation}
where $\mathbf{k}_{2d}=(k_x, k_y)$ and $\mathbf{r}_{2d}=(x, y)$. To solve the London equation Eq. (\ref{eq3}), we need to know the magnetic structure of the magnetic superconductor. In the following subsections, we treat the cases with magnetic moments perpendicular and parallel to the interface separately using the linear response approximation for the magnetization. 

\subsection{Magnetic moments parallel to the interface}
We assume the easy axis is along the $x$ direction. The magnetization can be written as $M_x=\chi B_x$, which is valid when $M_x\ll M_0$ with $M_0$ being the saturation magnetization. The solution to Eq. (\ref{eq3}) can be written as $\mathbf{B}=\mathbf{B}_3+\mathbf{B}_4$ with $\mathbf{B}_3$ accounting for the magnetic fields induced by the vortex and $\mathbf{B}_4$ being the solution to Eq. (\ref{eq3}) in the absence of vortices. Since the vortex is along the $z$ axis, we have $B_{3,x}=B_{3,y}=0$ and 
\begin{equation}\label{eq7}
B_{3,z}(k_x, k_y)=\frac{\Phi _0}{2\pi  (\mathbf{k}_{2d}^2\lambda _L^2+1)}.
\end{equation}
$\mathbf{B}_4$ in the Fourier space is given by
\begin{equation}\label{eq8}
\left(\mathbf{k}^2\lambda _L^2+1\right)B_{4,x}-4\pi  \lambda _L^2\left(k_y^2+k_z^2\right)\chi  B_{4,x}=0,
\end{equation}
\begin{equation}\label{eq9}
\left(\mathbf{k}^2\lambda _L^2+1\right)\left( {\begin{array}{cc}
 B_{4,y}  \\
 B_{4,z}
 \end{array} } \right)+4\pi  \lambda _L^2k_x\left( {\begin{array}{cc}
 k_y  \\
 k_z
 \end{array} } \right)\chi  B_{4,x}=0.
\end{equation}
From Eq. (\ref{eq8}), we obtain
\begin{equation}\label{eq10}
 k_z^2=\frac{4\pi  \chi  k_y^2-\lambda _L^{-2}-\mathbf{k}_{2d}^2}{1-4\pi  \chi }.
\end{equation}
From Eq. (\ref{eq9}), we obtain
\begin{equation}\label{eq11}
B_{4,y}=-\frac{k_xk_y}{k_y^2+k_z^2}B_{4,x} {\rm{, \ \ and\ \ }} B_{4,z}=-\frac{k_xk_z}{k_y^2+k_z^2} B_{4,x}.
\end{equation}
Using the boundary condition Eqs. (\ref{eq4}) and (\ref{eq5}), we have for $\mathbf{B}(\mathbf{k}_{2d}, z=0)$
\begin{equation}\label{eq12}
-\frac{1}{\sqrt{2\pi}}\frac{k_xk_z}{k_y^2+k_z^2} B_{4,x}+\frac{\Phi _0}{2\pi  (k_{2d}^2\lambda _L^2+1)}=B_{1,z}+ B_{2,z},
\end{equation}
\begin{equation}\label{eq13}
-\frac{1}{\sqrt{2\pi}}\frac{k_xk_y}{k_y^2+k_z^2} B_{4,x}= B_{1,y}+B_{2,y},
\end{equation}
\begin{equation}\label{eq14}
\frac{1-4\pi  \chi}{\sqrt{2\pi}}B_{4,x}=B_{1,x}+B_{2,x}.
\end{equation}
Using $\nabla\cdot \mathbf{B}_2=0$, we can derive the field $B_{4,x}$ from Eqs. (\ref{eq12}-\ref{eq14}). Substituting the results back into Eq. (\ref{eq12}), we have the magnetic fields outside the superconductor $B_{2,z}=B_{s,z}+B_{v,z}$, with the contribution from the magnetic source
\begin{equation}\label{eq15}
B_{s,z}(\mathbf{k}_{2d}, z=0)=-\alpha \left[\left(k_{2d}+\frac{1}{\alpha }\right)B_{1,z}-i k_xB_{1,x}-i k_yB_{1,y}\right],
\end{equation}
and the contribution from the vortex
\begin{equation}\label{eq16}
B_{v,z}(\mathbf{k}_{2d}, z=0)=\left(\alpha k_{2d}+1\right)\frac{\Phi _0}{2\pi(\mathbf{k}_{2d}^2\lambda _L^2+1)},
\end{equation}
with
\begin{equation}\label{eq17}
\alpha \left(k_x,k_y\right)=-k_z\left[k_z k_{2d}+i k_z^2-i 4\pi  \chi \left(k_y^2+k_z^2\right)\right]^{-1}.
\end{equation}
The magnetic field outside the superconductor then is given by
\begin{equation}\label{eq18}
B_{2,z}(\mathbf{r})=\int \frac{d^2 k_{2d}}{2\pi} B_{2,z}(\mathbf{k}_{2d}, z=0)\exp(-k_{2d}z+i \mathbf{k}_{2d}\cdot\mathbf{r}_{2d}).
\end{equation}
 Since the magnetic moments couple directly to the magnetic induction $\mathbf{B}$, we use the definition that $\chi=M_x/B_x$. In literatures, for examples see Refs. \onlinecite{Kirtley2012,Tachiki79,Gray83}, another definition $\chi'=M_x/H_x$ was used, where $\mathbf{H}=\mathbf{B}-4\pi \mathbf{M}$ is the external field "seen" by the magnetic moments. The relation between $\chi$ and $\chi'$ is $\chi'=\chi/(1-4\pi\chi)$. If $\chi'$ is introduced, one should replace $\sqrt{1-4\pi\chi}$ in the results of the present work by $1/\sqrt{1+4\pi\chi'}$. Please note that the magnetic susceptibility $\chi=M_x/B_x$ is smaller than $1/(4\pi)$, i.e. $\chi<1/(4\pi)$. The magnetic fluctuations $\left\langle M_x M_x\right\rangle\sim \chi/(1-4\pi \chi)$ diverges when $\chi\rightarrow 1/(4\pi)$, which indicates that the magnetic system becomes unstable \cite{Blount1979}.

\subsection{Magnetic moment perpendicular to the interface}
The calculations in this case are in parallel to those in the previous section. Here we skip the detailed calculations and only present the final results. The results can be obtained from Eqs. (\ref{eq15}) and (\ref{eq16}) by replacing $\alpha$ and $k_z$ with
\begin{equation}\label{eq19}
\alpha \left(k_x,k_y\right)=-\left(k_{2d}+i k_z\right)^{-1},
\end{equation}
\begin{equation}\label{eq20}
k_z^2=-\lambda _L^{-2} -\mathbf{k}_{2d}^2(1-4\pi \chi ).
\end{equation}

\section{Applying to MFM and SSM}
We have derived the general expressions for the magnetic field distribution $B_{2,z}$ outside the superconductor in response to the source field $B_{1,z}$. In the following subsection, we consider the cases of MFM and SSM respectively. For MFM, the magnetic source is modeled as a point dipole or monopole, and we then calculate the force between the  MFM tip and superconductors. For SSM, the source is modeled as a current loop and we calculate $B_{2,z}$. In both cases, the source magnetic field is extremely weak thus no additional vortex is induced by the source.

\subsection{Magnetic force microscopy}

In MFM, the force between the magnetic tip and the superconductor is measured as function of the distance between them\cite{Luan2010}. To calculate the force, one needs to know the magnetic field distribution inside the tip. Theoretical modeling of the tip is challenging since the magnetic field distribution and shape of the tip are generally unknown. In most treatments one assumes a single cylindrical magnetic domain with spatially uniformly distributed moments perpendicular to the sample surface.\cite{Hug1991,Bending1999} If the length of the cylinder is much larger than its radius, one can approximate the tip as a magnetic monopole. Otherwise the tip behaviors as a dipole. First we model the MFM tip by a point dipole along the $z$ direction
\begin{equation}\label{eq21}
\mathbf{M}_s=m_0 \delta \left(x\right)\delta \left(y\right)\delta (z-a)\hat{\mathbf{z}},
\end{equation}
where $a$ is the separation between the MFM tip and the superconductor. The approximation of the tip by a point dipole is valid when the size of the tip is much smaller than $a$. The typical size of the tip is tens of nanometer. For $a\gg \lambda_L$, it was shown that the shape of the MFM tip will not affect the results substantially.\cite{Xu1995} $\mathbf{B}_1$ then can be expressed as
\begin{equation}\label{eq22}
B_{1,z}(\mathbf{k}_{2d}, z=0)=m_0 \exp(-a k_{2d}) k_{2d},
\end{equation}
\begin{equation}\label{eq23}
\left( {\begin{array}{cc}
 B_{1,x}(\mathbf{k}_{2d}, z=0)  \\
 B_{1,y}(\mathbf{k}_{2d}, z=0)
 \end{array} } \right) =i m_0 \exp(-a k_{2d})\left( {\begin{array}{cc}
 k_x  \\
 k_y
 \end{array} } \right).
\end{equation}
The interaction between the tip and magnetic field is
\begin{eqnarray}\label{eq24}
\nonumber &&U(a)=-\int d^3r M_z (B_{1,z}+B_{2,z})\\
&&=-m_0[B_{1,z}(0, 0, a)+B_{2,z}(0, 0, a)].
\end{eqnarray}
The force then is given by $F=-\partial_a U(a)=F_{s}+F_{v}$ with the contribution from the source
\begin{equation}\label{eq25}
F_{s}=\frac{m_0^2}{\pi }\int \left(2\alpha k_{2d} + 1\right)\exp(-2a k_{2d})\mathbf{k}_{2d}^2 d^2  k_{2d},
\end{equation}
and the contribution from the vortex
\begin{equation}\label{eq26}
F_{v}=\frac{m_0\Phi _0}{(2\pi)^2 }\int \frac{\alpha\mathbf{k}_{2d}^2 + k_{2d}}{\mathbf{k}_{2d}^2\lambda _L^2+1}\exp\left(-ak_{2d}\right)d^2k_{2d}.
\end{equation}

Analytical expression for the force can be obtained when $a\gg \lambda_L$. In this case, only small $\mathbf{k}_{2d}$ contributes to the integration. For the magnetic moments parallel to the interface, we obtain
\begin{equation}\label{eq27}
F_{\parallel,s}=\frac{3m_0^2}{4\lambda_L^4}\left(\frac{\lambda_L^4}{a^4}-4\frac{\lambda _L^5}{a^5\sqrt{1-4\pi  \chi }}\right),
\end{equation}
\begin{equation}\label{eq28}
F_{\parallel,v}=\frac{m_0\Phi _0}{\lambda _L^3 \pi}\left(\frac{3\lambda _L^4}{a^4\sqrt{1-4\pi  \chi }}-\frac{\lambda _L^3}{a^3}\right).
\end{equation}

For the magnetic moments perpendicular to the interface, we have
\begin{equation}\label{eq29}
F_{\perp,s}=\frac{3m_0^2}{4\lambda_L^4}\left(\frac{\lambda_L^4}{a^4}-4\frac{\lambda _L^5}{a^5}\right),
\end{equation}
\begin{equation}\label{eq30}
F_{\perp,v}=\frac{m_0\Phi _0}{\lambda _L^3 \pi}\left(\frac{3\lambda _L^4}{a^4}-\frac{\lambda _L^3}{a^3}\right).
\end{equation}

\begin{figure}[t]
\psfig{figure=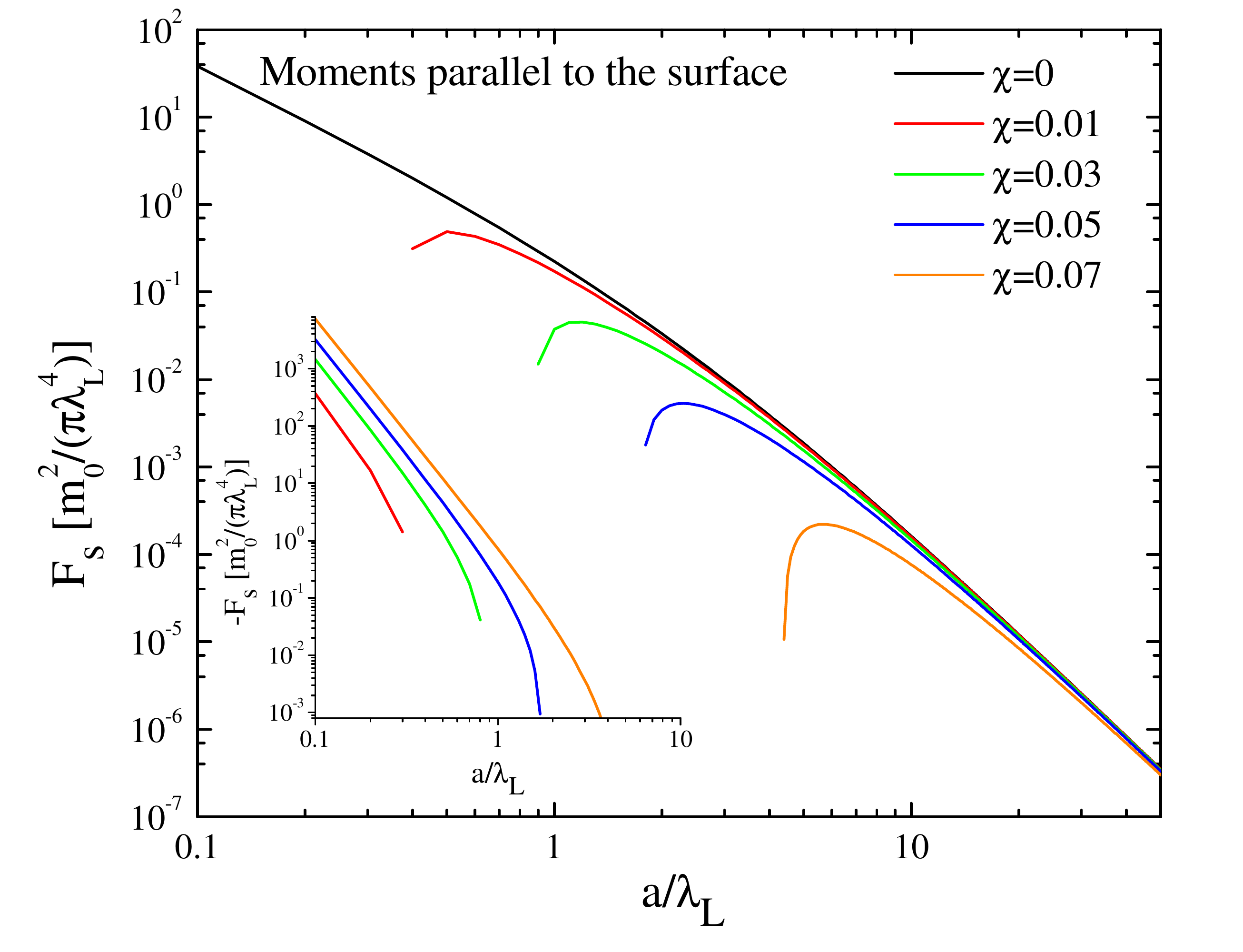,width=\columnwidth}
\caption{\label{f2}(color online) Force between a magnetic tip and magnetic superconductor in the absence of vortices obtained by numerical integration of Eq. (\ref{eq25}) when the magnetic moments are parallel to the surface. For small $a/\lambda_L$, the interaction becomes attractive for nonzero $\chi$ as shown in the inset. At large $a/\lambda_L\gg 1$, the curves are described by Eq. (\ref{eq27}).} 
\end{figure}
\begin{figure}[b]
\psfig{figure=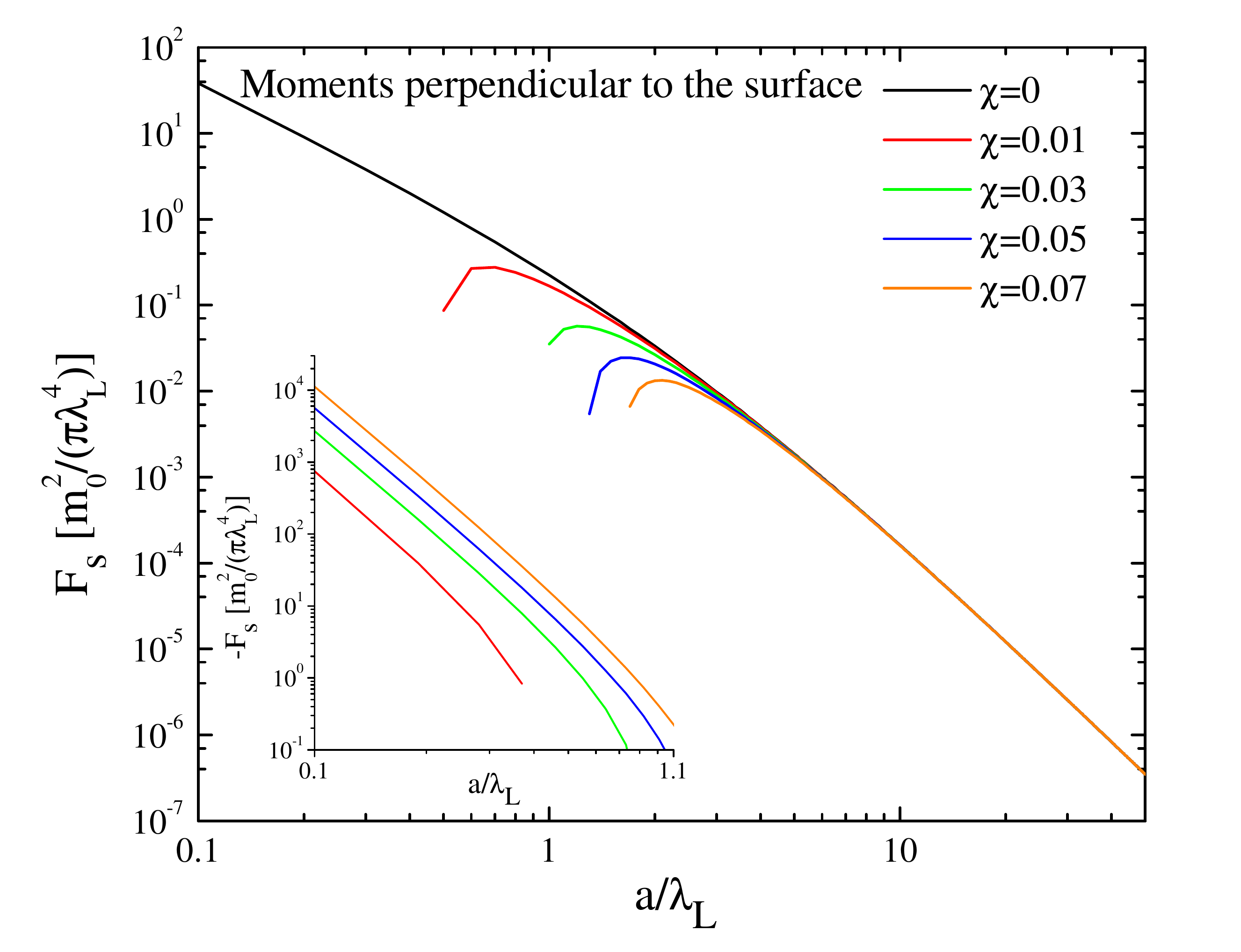,width=\columnwidth}
\caption{\label{f3}(color online) The same as Fig. \ref{f2} but for the magnetic moments perpendicular to the surface. Results are obtained by numerical integration of Eq. (\ref{eq25}).} 
\end{figure}

The exclusion of the magnetic flux by the superconductor gives rise to repulsion between the tip and superconductor, which is described by the first term in Eqs. (\ref{eq27}) and (\ref{eq29}). The force does not depends on the direction of the point dipole. For a magnetic superconductor, the polarization of magnetic moment reduces energy and causes attraction, as described by the second term in Eqs. (\ref{eq27}) and (\ref{eq29}). When the separation $a$ reduces, the attraction may be even larger than the repulsion, as shown by direct numerical integration of Eqs. (\ref{eq25}) in Figs. \ref{f2} and \ref{f3}. The attraction increases with $\chi$. 

The interaction between the vortex and tip depends on the direction of the dipole and it is attractive at large separation $a\gg \lambda_L$ when they are parallel. To visualize a vortex, one scans the tip in experiments, and the force depends on the vortex position relative to the tip. The position dependence of the force can be readily evaluated by replacing $\Phi_0\leftarrow \Phi_0 \exp[i (k_x x_v+k_y y_v]$. Here $r_v=(x_v, y_v)$ is the coordinate of the vortex core. For $a\gg \lambda_L$, the vortex-position-dependent force is 
\begin{equation}\label{eq31}
F(r_v=0)-F(r_v)=-\frac{3m_0\Phi _0}{\pi^2}\frac{ \lambda _L^2}{a^5} r_v^2.
\end{equation}

To extract the penetration depth in experiments, one measures the force as a function of $a$ in the absence of vortex.  One then obtains the penetration depth by fitting to theoretical expressions, such as Eq. (\ref{eq25}). In the magnetic superconductors, the force depends on the orientation of the magnetic moments with respect to the surface. For the magnetic moments parallel to the surface, one extracts an effective penetration depth $\lambda_L/\sqrt{1-4\pi\chi}$, see Eq. (\ref{eq27}). For the moments perpendicular to the surface, the bare penetration depth $\lambda_L$ is extracted, see Eq. (\ref{eq29}). By measuring the force in two different orientations, one can extract both $\chi$ and $\lambda_L$.

We proceed to model the magnetic tip as a monopole and calculate the interaction force. The magnetic field outside the superconductor is given by
 \begin{equation}\label{eqmo1}
\nabla\cdot\mathbf{B}=4\pi n_0 \delta(x)\delta(y)\delta(z-a), \ \ \ \nabla\times\mathbf{B}=0.
\end{equation}
Here $n_0$ is a magnetic charge. The magnetic field distribution outside $B_{2,z}$ can still be calculated with Eqs. (\ref{eq15}) and (\ref{eq18}) with $\alpha$ and $k_z$ given by Eqs. (\ref{eq10}), (\ref{eq17}), (\ref{eq19}), (\ref{eq20}) depending on the orientation of the magnetic moments inside the superconductor. $\mathbf{B}_1$ in this case is given by
\begin{equation}\label{eqmo2}
B_{1,z}(\mathbf{k}_{2d}, z=0)=-n_0 \exp(-a k_{2d}),
\end{equation}
\begin{equation}\label{eqmo3}
\left( {\begin{array}{cc}
 B_{1,x}(\mathbf{k}_{2d}, z=0)  \\
 B_{1,y}(\mathbf{k}_{2d}, z=0)
 \end{array} } \right) =-i n_0 \frac{\exp(-a k_{2d})}{k_{2d}}\left( {\begin{array}{cc}
 k_x  \\
 k_y
 \end{array} } \right).
\end{equation}
The $z$-component force is given by $F=n_0 B_{2,z}(0, 0, a)$. When the magnetic moments are parallel to the surface, we obtain the force due to the source $F_{\parallel,s}$ and the force due to vortex $F_{\parallel,v}$ in the limit $a\gg\lambda_L$
\begin{equation}\label{eqmo4}
F_{\parallel,s}=\frac{n_0^2}{ \lambda _L^2}\left(\frac{\lambda _L^2}{4 a^2}-\frac{\lambda _L^3}{2 a^3\sqrt{1-4\pi  \chi }}\right),
\end{equation}
\begin{equation}\label{eqmo5}
F_{\parallel,v}=\frac{n_0\Phi _0}{2\pi  \lambda _L^2}\left(\frac{\lambda _L^2}{a^2}-\frac{4\lambda _L^3}{a^3\sqrt{1-4\pi  \chi }}\right).
\end{equation}
For the magnetic moments perpendicular to the surface, we have
\begin{equation}\label{eqmo6}
F_{\perp,s}=\frac{n_0^2}{ \lambda _L^2}\left(\frac{\lambda _L^2}{4 a^2}-\frac{\lambda _L^3}{2 a^3}\right),
\end{equation}
\begin{equation}\label{eqmo7}
F_{\perp,v}=\frac{n_0\Phi _0}{2\pi  \lambda _L^2}\left(\frac{\lambda _L^2}{a^2}-\frac{4\lambda _L^3}{a^3}\right).
\end{equation}
In the limit $\chi\rightarrow 0$ and $\lambda_L\rightarrow 0$, one can model the superconductor as a perfect magnetic conductor and the magnetic field outside the superconductor can be obtained with the image method. If one models the MFM tip as a magnetic dipole, the repulsion between the image dipole and tip is ${3 m_0^2}/(4 a^4)$. If the tip is treated as a monopole, the repulsion force is $n_0^2/(4a^2)$. Both Eqs. (\ref{eq27}) and (\ref{eqmo4}) reproduce the limiting results.

\subsection{Scanning SQUID microscopy}
In SSM, one applies external magnetic field through a field coil and then measures the magnetic field above the superconductor through a pickup loop \cite{Kirtley2010,Kirtley2012}. We model the field coil by a loop with  current $I$ and radius $r_0$. Please note that when $a\gg r_0$, the magnetic field induced by the current loop is reduced to the point dipole discussed in the previous section. The source magnetic field due to the current loop is then given by $\mathbf{B}_1=\nabla\varphi$ with the magnetic potential\cite{Kogan2003}
 \begin{equation}\label{eq32}
\varphi (r)= \int {d^2 k_{2d}} \varphi (\mathbf{k}_{2d}) \exp \left[i \mathbf{k}_{2d}\cdot\mathbf{r}_{2d}+k_{2d}z\right],
\end{equation}
 \begin{equation}\label{eq33}
\varphi \left(\mathbf{k}_{2d}\right)= \frac{r_0 I}{c k_{2d}}\exp \left(-k_{2d}a\right) J_1\left(k_{2d} r_0\right).
\end{equation}
In this case, the magnetic field outside the superconductor due to the source field is 
\begin{equation}\label{eq34}
B_{s,z}(\mathbf{k}_{2d}, z=0)=-\alpha \left(2k_{2d}+\frac{1}{\alpha }\right)\frac{2\pi I a}{c}J_1\left(k_{2d} r_0\right)\exp \left(-k_{2d} a\right)
\end{equation}
and the vortex contribution is the same as Eq. (\ref{eq16}). Here $\alpha$ depends on the orientation of the magnetic moments. For the moments parallel to the surface, $\alpha$ is given by Eq. (\ref{eq17}) and for the moments perpendicular to the surface it is given by Eq. (\ref{eq19}). When the magnetic moments are parallel to the surface, the magnetic field at the center of the pickup loop is given by for $a\gg \lambda_L$
\begin{equation}\label{eq35}
B_{s,z}=\frac{\pi  I r_0^2}{4c a^3}\left(-1+\frac{3 \lambda _L}{ a\sqrt{1-4 \pi  \chi }}\right),
\end{equation}
\begin{equation}\label{eq36}
B_{v,z}=\frac{\Phi _0}{2\pi a^2}\left(1-\frac{2 \lambda _L}{a\sqrt{1-4 \pi  \chi }}\right).
\end{equation}
For the moments perpendicular to the surface, the results are the same as those in Eqs. (\ref{eq35}) and (\ref{eq36}), but without the factor $\sqrt{1-4\pi\chi}$, similar to the case of MFM.

In the case of a magnetic superconductor with isotropic magnetic structure as studied in Ref. \onlinecite{Kirtley2012}, the magnetic field outside the superconductor is given by Eqs. (\ref{eq35}) and (\ref{eq36}). The extracted penetration depth from SSM measurements is $\lambda _L/\sqrt{1-4 \pi  \chi }$ which is larger than the bare $\lambda_L$. This is different from the effective penetration depth in magnetic superconductors $\lambda_L\sqrt{1-4 \pi  \chi }$, which is smaller than the bare $\lambda_L$.

\section{Conclusion}
We have calculated the magnetic fields outside a magnetic superconductor when a magnetic source is placed on top of the superconductor. For the magnetic moments parallel to the surface, the resulting magnetic field distribution depends on an effective penetration depth $\lambda_L/\sqrt{1-4\pi\chi}$  when the distance between the magnetic source and superconductor is much larger than $\lambda_L$, while for the moments perpendicular to the surface, it depends on $\lambda_L$. The results in the present work can be used to measure both the susceptibility $\chi$ and the penetration depth $\lambda_L$ in magnetic superconductor by the magnetic force microscopy or scanning SQUID microscopy. This can be achieved by changing the orientation of the crystal.

\section{Acknowledgment}
The authors thank Jeehoon Kim and Matthias J. Graf for helpful discussions. This work was supported by the US Department of Energy, Office of Basic Energy Sciences, Division of Materials Sciences and Engineering.

%

\end{document}